\documentclass[a4paper,aps,prd,onecolumn,showpacs,showkeys,nofootinbib,preprintnumbers,superscriptaddress,amsmath,amssymb,amsfonts]{revtex4}
\usepackage{graphicx}
\usepackage{bm,natbib}
\usepackage{amsmath}
\usepackage[english]{babel}
\usepackage[T1]{fontenc}
\usepackage{amssymb}
\usepackage{amsfonts}

\usepackage{amsthm}              

\usepackage{braket}
\usepackage{quoting}
\quotingsetup{font=small}
\usepackage{csquotes}
\usepackage{url}

\usepackage{epsfig}
\usepackage{colordvi}
\usepackage{psfrag}
\usepackage{color}
\usepackage{mathrsfs}

\newcommand{\Lagr}{\mathcal{L}}

\usepackage{dcolumn}
\usepackage{multirow}
\usepackage{hyperref}
\usepackage{epstopdf}
\usepackage{bm}
\usepackage{times}
\graphicspath{ {Bouncing f(Q) and Wave Function/} }

\hypersetup{
                    colorlinks=true,        
                    linkcolor=blue,         
                    citecolor=magenta,      
}

\begin{document}
\title{Bouncing Cosmology in $f(Q)$ Symmetric Teleparallel Gravity}

\author{Francesco Bajardi}
\email{bajardi@na.infn.it}
\affiliation{Department of Physics ``E. Pancini'', University of Naples ``Federico II'', Naples, Italy.}
\affiliation{INFN Sez. di Napoli, Compl. Univ. di Monte S. Angelo, Edificio G, Via Cinthia, I-80126, Naples, Italy.}

\author{Daniele Vernieri}
\email{daniele.vernieri@unina.it}
\affiliation{Department of Physics ``E. Pancini'', University of Naples ``Federico II'', Naples, Italy.}
\affiliation{INFN Sez. di Napoli, Compl. Univ. di Monte S. Angelo, Edificio G, Via	Cinthia, I-80126, Naples, Italy.}

\author{Salvatore Capozziello}
\email{capozziello@na.infn.it}
\affiliation{Department of Physics ``E. Pancini'', University of Naples ``Federico II'', Naples, Italy.}
\affiliation{INFN Sez. di Napoli, Compl. Univ. di Monte S. Angelo, Edificio G, Via	Cinthia, I-80126, Naples, Italy.}
\affiliation{Scuola Superiore Meridionale, Largo San Marcellino 10, I-80138, Naples, Italy.}
\affiliation{Tomsk State Pedagogical University, ul. Kievskaya, 60, 634061 Tomsk, Russia.}

\date{\today}

\begin{abstract}
We consider $f(Q)$ extended symmetric teleparallel cosmologies, where $Q$ is the non-metricity scalar, and constrain its functional form through the order reduction method. By using this technique, we are able to reduce and integrate the field equations and thus to select  the corresponding models giving rise to bouncing cosmology. The selected Lagrangian is then used to develop the Hamiltonian formalism and to obtain the Wave Function of the Universe which suggests that classical observable universes can be recovered according to the Hartle Criterion. 
\end{abstract}


\maketitle

\section{Introduction}

The gravitational interaction, described by Einstein's General Relativity (GR), is the only fundamental force escaping a formulation according to Quantum Field Theory. After many attempts, the difficulty of quantizing gravity arose for several reasons. For instance, there are no techniques able to delete the divergences occurring in the two-loop effective action, so that the theory turns out to be renormalizable just up to one-loop level. In any case, there is also a lack in the quantum side because, in view of Quantum Gravity, the metric tensor should act both as a fundamental field and as the background. This makes the construction of a theory of Quantum Gravity very difficult starting from fundamental concepts.

On the other hand, as widely demonstrated during the last decades, quantum corrections play a crucial role at infrared and ultraviolet scales, providing a fundamental contribution toward the explanation of late and early Universe behavior. 

For example, the Big Bang theory suffers the initial singularity problem: spacetime should enucleate from ``nothing'' with deep conceptual issues related to this statement. Despite the lack of a final theory of  Quantum Gravity, we can still fix some issues by considering the applications to cosmology. The approach consists in deriving dynamical quantum systems related to cosmological models and testable, in principle, by means of observations. This is not the full Quantum Gravity, but it is a workable scheme towards it. For example, from the Loop Quantum Cosmology (LQC), it is possible to get bouncing solutions, according to which the Universe might cyclically undergo an accelerated expansion followed by a contraction~\cite{Bojowald:2006da, Singh:2006im, Singh:2006sg, WilsonEwing:2012pu, Cai:2014zga, Mielczarek:2008zv}. Specifically, the \emph{Big Bounce} theory predicts an exponential expansion in the Universe early stage, ending with an accelerated collapse (\emph{Big Crunch}). 

High-energy issues are not the only shortcomings suffered by GR: the theory is even not capable of explaining dark components which constitute the 95 $\%$ of the bulk of the Universe at large scales.
Solution for this issue might be found in extensions and modifications of GR action, according to which dark energy and dark matter can be addressed as curvature effects at astrophysical and cosmological scales~\cite{ Capozziello:2012ie, Capozziello:2003tk, Nojiri:2010wj, Nojiri:2017ncd, Linder:2010py, Nojiri:2005vv}. One of the most renowned proposal in this sense is represented by $f(R)$ gravity~\cite{Capozziello:2002rd, Sotiriou:2008rp, DeFelice:2010aj, Cognola:2007zu}, where the gravitational action of GR is extended including a generic function of the scalar curvature. Indeed, by means of such an extension, it is possible to explain the current acceleration of the Universe~\cite{Koyama:2015vza, Capozziello:2011et} and to fit the galaxy rotation curves~\cite{Famaey:2011kh, Capozziello:2006ph, Stabile:2013jon, Jovanovic:2018jtz, Capozziello:2017rvz} without any dark energy or dark matter. $f(R)$ gravity is just an example of more general classes of theories leading to higher-order field equations (see \emph{e.g.} Refs.~\cite{Capozziello:2011et, Clifton:2011jh, Capozziello:2012hm, deRham:2014zqa, Jackiw:2003pm, Capozziello:2019klx, Bajardi:2020osh, Bajardi:2019zzs, Bajardi:2020xfj, Vernieri:2011aa, Vernieri:2012ms, Vernieri:2015uma, Frusciante:2015maa} for other modified and extended  theories of gravity). 

Another assumption of GR is to use the Levi--Civita connection. It is required because the Equivalence Principle is the foundation of Einstein's theory by which the geodesic and the metric structures of spacetime coincide. As a consequence, isometries and universality of free fall are preserved and spacetime is torsionless.

Once the hypotheses of torsionless and metric-compatible connection are relaxed, indeed, two other theories, equivalent to GR, can be constructed. In particular, assuming the antisymmetric part of the connection $\Gamma^\alpha_{\,\,\,\, \mu \nu}$ to be different from zero, a torsion contribution in the spacetime arises and gravity is described by the so called \emph{Einstein--Cartan} theory~\cite{Hehl:1994ue, Plebanski:1977zz}. 

Moreover, by setting the curvature to zero, the spacetime is only described by the torsion and the resulting theory is called \emph{Teleparallel Equivalent to General Relativity} (TEGR). In this case, a more general affine structure has to be considered, the so called Weitzenb\"ock connection~\cite{Arcos:2005ec, Hammond:2002rm, Maluf:2013gaa}. It is formulated by tetrad fields on the tangent space and then TEGR can be recast as a theory for the translation group in the local tangent spacetime. In any case the teleparallel action is equivalent to the GR one up to a boundary term, so that the field equations are exactly the same. As a consequence of this equivalence, neither TEGR nor GR are able to solve the above mentioned problems of the large-scale structure; in analogy to modified metric theories of gravity such as $f(R)$ theories, modified teleparallel actions aims to solve the cosmological and astrophysical issues by introducing  functions of the torsion scalar $T$ or other second-order torsion invariants~\cite{Cai:2015emx, Li:2010cg, Wu:2010mn, Benetti:2020hxp, Capozziello:2019msc}. The advantage of dealing with $f(T)$ models instead of $f(R)$ ones, is due to the order of the field equations; while $f(R)$ gravity leads to fourth-order field equations in metric formalism, in $f(T)$ gravity the corresponding equations are of second order. This allows to simplify the dynamics and to find easily exact solutions. 

Another class of theories whose Christoffel connection is different from the Levi--Civita connection, are the so-called \emph{Non-Metric Theories}, according to which the covariant derivative of the metric tensor does not vanish identically and a new tensor quantity can be constructed, \emph{i.e.}, $Q_{\mu \alpha\beta} \equiv \nabla_{\mu} g_{\alpha \beta} \neq 0$. These theories do not require the validity of Equivalence Principle at the fundamental level.

After defining a \emph{non-metricity scalar} $Q$, the action $S = (\kappa/2) \int \sqrt{-g} \,  Q \, d^4x$ turns out to be the same as that of TEGR and then  the corresponding theory  is called \emph{Symmetric Teleparallel Equivalent to General Relativity} (STEGR). The field equations are those of the  Einstein theory and then, in this sense, GR, TEGR and STEGR are equivalent and give rise to the so-called \emph{Gravity Trinity}~\cite{BeltranJimenez:2019tjy}.
In the action above, $\kappa = 1/(8 \pi G_N)$ is the gravitational coupling, $G_N$ is the Newton's gravitational constant and $g$ is the determinant of the metric $g_{\mu \nu}$. In analogy to $f(T)$ and $f(R)$ gravity, modified non-metric theories with action $S = -(1/2) \int \sqrt{-g} \, f(Q) \, d^4 x$ can be considered; however, even if STEGR, TEGR and GR are  interchangeable, their extensions are different from each other: while $f(Q)$ is equivalent to $f(T)$, $f(R)$ gravity leads to a different dynamics.
 
In view of this difference, it is particularly useful to study related cosmologies with the aim to reconstruct cosmic histories capable of matching large datasets at any epoch and then select a self-consistent theory of gravity. In particular, being Quantum Cosmology related to the law of initial conditions from which the observed Universe emerged, considering wide classes of models can be a useful approach to avoid any fine-tuning issue. 

In this paper we study bouncing cosmology in $f(Q)$ gravity and we select bouncing solutions by means of the order reduction method of the field equations~\cite{Bel:1985zz,Simon:1990ic,Simon:1991bm}. Since modified theories of gravity carry out further degrees of freedom, the field equations are often hardly solvable and, for this reason, the order reduction is a useful approach to solve dynamics. The same approach has been adopted \emph{e.g.} in Refs.~\cite{Sotiriou:2008ya, Terrucha:2019jpm, Barros:2019pvc}, where the authors find the form of the actions in agreement with the Big Bounce theory. In order to develop Quantum Cosmology in $f(Q)$ gravity, we take into account the \emph{Arnowitt--Deser--Misner} (ADM) formalism in modified STEGR action. The Hamiltonian formalism and the related quantization permits to find out the \emph{Wave Function of the Universe}, whose behavior gives information on the possibility to realize observable universes. Indeed, according to the \emph{Hartle Criterion}, the Wave Function describes correlations among cosmological observables: if it oscillates, cosmological parameters are correlated and then can give rise to observables universes, whose dynamics is described by classical trajectories. In this perspective, studying generalized STEGR models is useful since non-metricity can enlarge the set of viable minisuperspaces.

The layout of the paper is the following. Sec.~\ref{STEGR} is devoted to discuss the main features of STEGR and its modifications, defining all the quantity needed to construct the modified non-metric dynamics. In Sec.~\ref{OrderRedSect}, after recalling the general technique, we search for $f(Q)$ models leading to bouncing cosmology. In Sec.~\ref{WFsect}, the ADM formalism is applied to the selected function of the non-metricity scalar and, hence, a solution of the Wheeler--DeWitt equation is studied. Finally, in Sec.~\ref{concl} we conclude this work summing up the main results and discussing the future perspectives.
The Hamiltonian formulation of GR and the ADM formalism are summarized in Appendix~\ref{HAM}.

\section{Modified Non-Metric Theories of Gravity} \label{STEGR}

As it is well known, in non-flat spacetimes, geodesic structure is assigned by the form of the connection. In GR, the assumption of torsionless and metric-compatible connection gives the Levi--Civita connection, related to the metric and its first derivatives; once relaxing such an hypothesis, it is possible to define two rank-3 tensors linked to the antisymmetric part of $\Gamma^\rho_{\,\,\,\, \mu \nu}$ and to the covariant derivative of the metric:
\begin{equation}
T_{\,\,\,\, \mu \nu}^{\alpha}=2 \Gamma_{\,\,\,\, [\mu \nu]}^{\alpha}\,, \quad Q_{\rho \mu \nu} \equiv \nabla_{\rho} g_{\mu \nu} \neq 0\,.
\end{equation}
The former is called \emph{torsion tensor}, while the latter is called \emph{non-metricity tensor}. It follows that the most general connection comprehending all possible contributions read as
\begin{equation}
\Gamma^{\rho}_{\,\,\,\,\mu\nu}=\breve{\Gamma}_{\,\,\,\, \mu \nu}^{\rho}+K^{\rho}_{\,\,\,\,\mu\nu}+L^{\rho}_{\,\,\,\,\mu\nu}\,,
\end{equation}
where $\breve{\Gamma}_{\,\,\,\, \mu \nu}^{\rho}$ is the Levi--Civita connection,
\begin{equation}
K^{\rho}_{\,\,\,\,\mu\nu}\equiv\frac{1}{2}g^{\rho\lambda}\bigl(T_{\mu\lambda\nu}+T_{\nu\lambda\mu}+T_{\lambda\mu\nu}\bigr)%
                                      =-K^{\rho}_{\,\,\,\,\nu\mu} \,,
\end{equation}
and
\begin{equation}
L^{\rho}_{\,\,\,\,\mu\nu}\equiv\frac{1}{2}g^{\rho\lambda}\bigl(-Q_{\mu \nu \lambda}-Q_{\nu \mu \lambda} + Q_{\lambda\mu\nu}\bigr)=%
                                               L^{\rho}_{\,\,\,\,\nu\mu}\,.
\end{equation}
GR assumes that both the  \emph{Contorsion Tensor}  $K^{\rho}_{\,\,\,\,\mu\nu}$  and the \emph{Disformation Tensor}
$L^{\rho}_{\,\,\,\,\mu\nu}$ identically vanish. Depending on the form of the connection, one can construct three different theories, namely:
\begin{equation}
    \begin{array}{l} \mbox{GR} \rightarrow L^{\rho}{ }_{\mu \nu}=K^{\rho}{ }_{\mu \nu}=0\,, \\ \mbox{TEGR} \rightarrow \breve{\Gamma}^{\rho}{ }_{\mu \nu} =L^{\rho}{ }_{\mu \nu}=0\,, \\ \mbox{STEGR} \rightarrow \breve{\Gamma}^{\rho}{ }_{\mu \nu}=K^{\rho}{ }_{\mu \nu}=0\,.\end{array}
\end{equation}
GR describes the spacetime through the curvature, setting to zero the torsion and the non-metricity; TEGR dynamics is given by  torsion while curvature and non-metricity vanish; in STEGR only  non-metricity describes dynamics. In any case, the three theories are completely equivalent at the level of field equations and the choice depends on what variables are assumed to describe the gravitational interaction. The assumption of non-metricity to label geometry, implies that a given vector changes its norm while parallel transported along the spacetime. In the same way, the torsion leads to a shift of the vector after performing a closed path. By means of the definitions
\begin{eqnarray}
&& S_{p \mu \nu}=-S^{p \mu \nu}=K^{\mu \nu p}-g^{p \nu} T_{\,\,\,\,\,\,\, \sigma}^{\sigma \mu}+g^{p \mu} T_{\,\,\,\,\,\,\, \sigma}^{\sigma \nu}\,,
\\
&& Q \equiv - \frac{1}{4} Q_{\alpha \mu \nu} \left[- 2 L^{\alpha \mu \nu}+  g^{\mu \nu} \left(Q^\alpha - \tilde{Q}^\alpha \right) - \frac{1}{2} \left(g^{\alpha \mu} Q^\nu + g^{\alpha \nu} Q^\mu  \right)\right],
\\
&& Q_\mu \equiv Q_{\mu \,\,\,\, \lambda}^{\,\,\,\lambda}\,,
\\
&& \tilde{Q}_{\mu}=Q_{\alpha \mu}^{\, \, \, \, \, \, \alpha}\,,
\end{eqnarray}
it turns out that the three actions 
\begin{eqnarray}
&&\mathcal{S}_{GR}\equiv \frac{\kappa}{2}  \int d^4x\,\sqrt{-g}\, R +\mathcal{S}^{(m)}\,,\\
&&\mathcal{S}_{TEGR}\equiv \frac{\kappa}{2} \int d^4x\,\sqrt{-g}\,T+\mathcal{S}^{(m)}\,,\label{TGACT}\\ 
&&\mathcal{S}_{STEGR}\equiv \frac{\kappa}{2} \int d^4x\,\sqrt{-g}\,Q+\mathcal{S}^{(m)}\,, \label{stegraction}
\end{eqnarray}
differ from each other only by a four-divergence.
We do not further investigate TEGR here; for details on possible applications see, e.g.  Refs.~\cite{Cai:2015emx, Capozziello:2011hj,Boehmer:2011gw, Wu:2010xk, Finch:2018gkh}. In what follows, we  focus on STEGR and on a modified action containing a function of the non-metricity scalar $Q$. By varying the action in Eq.~\eqref{stegraction}, we get the field equations~\cite{Sotiriou:2009xt,BeltranJimenez:2019tjy}
\begin{eqnarray}
&& \frac{2}{\sqrt{-g}} \nabla_{\alpha}\left\{\sqrt{-g} g_{\beta \nu} \left[L^{\alpha \mu \beta}- \frac{1}{2} g^{\mu \beta} \left(Q^\alpha -  \tilde{Q}^\alpha \right) + \frac{1}{4} \left(g^{\alpha \mu} Q^\beta + g^{\alpha \beta} Q^\mu  \right)\right]\right\}- \delta_{\nu}^{\mu} Q+ \nonumber \\
&&  \left[ L^{\mu \alpha \beta}- \frac{1}{2} g^{\alpha \beta} \left(Q^\mu -  \tilde{Q}^\mu \right) + \frac{1}{4} \left(g^{\mu \alpha} Q^\beta + g^{\mu \beta} Q^\alpha  \right)\right] Q_{\nu \alpha \beta}=T_{\,\,\,\nu}^{\mu}\,,
\end{eqnarray}
where $\nabla_{\alpha}$ denotes the covariant derivative with respect to the connection $L^{\rho}_{\,\,\,\,\mu\nu}$. The extension of the action to a function of $Q$, namely
\begin{equation}
    S = - \frac{1}{2}\int \sqrt{-g} \, f(Q) \, d^4 x\,,
    \label{f(Q)action}
\end{equation}
once varied with respect to the metric tensor, provides the field equations~\cite{Jimenez:2019ovq, Dialektopoulos:2019mtr}: 
\begin{eqnarray}
&& \frac{2}{\sqrt{-g}} \nabla_{\alpha}\left(\sqrt{-g} g_{\beta \nu} f_{Q} \left[- \frac{1}{2} L^{\alpha \mu \beta}+ \frac{1}{4} g^{\mu \beta} \left(Q^\alpha -  \tilde{Q}^\alpha \right) - \frac{1}{8} \left(g^{\alpha \mu} Q^\beta + g^{\alpha \beta} Q^\mu  \right)\right]\right)+\frac{1}{2} \delta_{\nu}^{\mu} f+ \nonumber \\
&& f_{Q} \left[- \frac{1}{2} L^{\mu \alpha \beta}+ \frac{1}{4} g^{\alpha \beta} \left(Q^\mu -  \tilde{Q}^\mu \right) - \frac{1}{8} \left(g^{\mu \alpha} Q^\beta + g^{\mu \beta} Q^\alpha  \right)\right] Q_{\nu \alpha \beta}=T_{\,\,\,\nu}^{\mu}\,.
\label{FE f(Q)}
\end{eqnarray}
 In next Sections, we will use the field equations~\eqref{FE f(Q)} in order to select the functional form of the gravitational action~\eqref{f(Q)action} admitting bouncing solutions; finally, we will study the corresponding Wave Function of the Universe.

\section{Bouncing Cosmology via Order Reduction of the Field Equations} \label{OrderRedSect}

Bouncing solutions play an important role in  cosmology, since they allow to avoid the Big Bang initial singularity describing a cyclic expansion  of the Universe. Any expansion is followed by a consequent contraction. Theory is based on the results provided by the application of Loop Quantum Gravity to Cosmology, and predicts a periodic succession of Big Bang and Big Crunch phases. In the low-energy limit, LQC agrees with Einstein GR, while, in the high-energy regime, it aims to solve the initial singularity problem by adopting a quantum description. 

Let us consider the spatially flat Friedmann--Lema\^itre--Robertson--Walker (FLRW) line element with the lapse function $N(t)$, and scale factor $a(t)$ (see Appendix \ref{HAM}) that is:
\begin{equation}
ds^2 = - N(t)^2 dt^2 + a(t)^2 \delta_{ij} dx^i dx^j\,.
\label{interval}
\end{equation} 
The non-metricity scalar reads~\cite{Dialektopoulos:2019mtr}:
\begin{equation}
Q = 6 \frac{H^2}{N^2}\,,
\label{Q(H)}
\end{equation}
where $H \equiv \dot{a}^2/a^2$. Notice that the second derivative of the scale factor does not appear in the non-metricity scalar expression and, as we will see, this leads to a Lagrangian which is independent of the time-derivative of $Q$.

Using the above metric, our purpose is to find bouncing cosmological solutions for $f(Q)$ gravity, whose action is given by Eq.~\eqref{f(Q)action}.
However, as it often happens in modified theories of gravity, the resulting field equations (in presence of matter) turn out to be analytically unsolvable. Then, we resort to the order reduction method of the field equations~\cite{Bel:1985zz,Simon:1990ic,Simon:1991bm}, by means of which we can write the geometric quantities in terms of matter fields at the lowest order in the perturbative expansion, selecting only those solutions which are perturbatively close to GR. 

Setting $N=1$, the first Friedmann equation reads~\cite{Jimenez:2019ovq}:
\begin{equation}
6 f_Q(Q) H^2 - \frac{1}{2} f(Q) = \rho\,.
\label{fried1}
\end{equation}
In order to implement the order reduction method, let us parametrize the function $f(Q)$ as:
\begin{equation}
f(Q) = \kappa Q + \varepsilon \varphi(Q)\,,
\label{varphi}
\end{equation}
where $\varepsilon \varphi(Q) \ll \kappa Q$, and the parameter $\varepsilon$ controls the deviation from STEGR.
In this way, as soon as $\varepsilon \to 0$, we recover the standard STEGR and, therefore, the field equations as in GR.

By replacing Eq.~\eqref{varphi} into Eq.~\eqref{fried1}, the first Friedmann equation becomes
\begin{equation}
3 \kappa H^2 + 6 \varepsilon \,\varphi_Q(Q) H^2 - \frac{1}{2} \varepsilon \,\varphi(Q) = \rho\,.
\label{friedmod}
\end{equation}
At the lowest order, when $\varepsilon \to 0$, the function $f$ is simply equal to the non-metricity scalar $(f = \kappa Q)$, and Eq.~\eqref{friedmod} provides
\begin{equation} \label{relation}
 \rho = \frac{\kappa}{2} Q\,,
\end{equation}
which is nothing but the standard Friedman equation for STEGR, written in terms of the non-metricity scalar. Then, Eq.~\eqref{friedmod} can be rewritten as 
\begin{equation} \label{comp1}
H^2 = \frac{\rho}{3 \kappa} + \frac{\varepsilon}{3 \kappa} \left( - 6 \varphi_Q H^2 + \frac{1}{2} \varphi \right), 
\end{equation}
so that it can be compared to the LQC equation in a cosmological background~\cite{Singh:2006sg}:
\begin{equation} \label{comp2}
H^{2}=\frac{\rho}{3 \kappa} \left(1-\frac{\rho}{\rho_{c}}\right). 
\end{equation}
The quantity $\rho_c$ is the critical energy density defined as $\rho_c = \sqrt{3}/(32 \pi^2 \gamma^3 G_N l_{Pl}^2)$, where $\gamma$ is the Barbero--Immirzi parameter, and $l_{Pl}$ is the Planck length.

The comparison between Eq.~\eqref{comp1} and Eq.~\eqref{comp2}, using Eq.~\eqref{relation}, yields:
\begin{equation}
- Q \varphi_Q + \frac{1}{2} \varphi + \frac{\kappa^2 Q^2}{4 \varepsilon \rho_c} = 0\,,
\end{equation}
whose solution is
\begin{equation}
\varphi(Q) = \frac{\kappa^2 Q^2}{6 \varepsilon \rho_c} + c_1 \sqrt{Q}\,,
\label{solf(Q)}
\end{equation}
where $c_1$ is an integration constant. The complete function leading to cosmological bouncing solutions, therefore, turns out to be
\begin{equation}
    f(Q) = \kappa Q + \frac{\kappa^2 Q^2}{6 \rho_c} + \tilde{c}_1 \sqrt{Q}\,,
    \label{solf(Q)1}
\end{equation}
with the definition $\tilde{c}_1 \equiv \varepsilon c_1$.

\section{Quantum Cosmology in $f(Q)$ Gravity and Wave Function of the Universe} \label{WFsect}

In this section, taking the previously calculated function $f(Q)$ which provides a bouncing cosmological model, we use the Hamiltonian formalism to obtain the Wheeler--DeWitt equation, whose solution gives the Wave Function of the Universe (see Appendix \ref{HAM}). We develop a general Lagrangian formalism for any $f(Q)$ actions, to subsequently focus on Eq.~\eqref{solf(Q)1} for the application of ADM formalism. Let us obtain cosmological point-like Lagrangian  for  the modified non-metric action~\eqref{f(Q)action}.  It can be reduced by integrating the three-dimensional surface and, thanks to Lagrange Multipliers Method, the point-like Lagrangian can be found. 
Using Eq.~\eqref{Q(H)}, the action in Eq.~\eqref{f(Q)action} can be written as:
\begin{equation}
S = - \frac{1}{2}\int \left\{a^3 N f(Q) - \lambda\left[Q - \frac{6}{N^2}\left(\frac{\dot{a}^2}{a^2}\right)\right] \right\}dt\,.
\end{equation}
The variation of the action with respect to the non-metricity scalar, yields the Lagrange multiplier $\lambda=a^3 N f_Q$. Finally, the cosmological point-like Lagrangian can be written as:
\begin{equation}
\Lagr = -\frac{3 a \dot{a}^2 f_Q}{N} - \frac{a^3 N}{2} (f - Q f_Q)\,.
\end{equation}
The minisuperspace considered is a three-dimensional space depending on the variables ${\cal{S}} = \{a,N,Q\}$. Thus, since the Lagrangian is independent of $\dot{Q}$ and $\dot{N}$, the tangent space is not a six-dimensional space but  contains only four variables: ${\cal{TS}} = \{a,N,Q,\dot{a}\}$. The Euler--Lagrange equations provide the following system of differential equations:
\begin{equation}
\begin{cases}
\displaystyle a: \,\, \, \, 2 N \dot{a}^2 f_Q - a^2 N^3 (f - Q f_Q ) +  4 a [N \ddot{a} f_Q + \dot{a} (-f_Q \dot{N} + N \dot{Q} f_{QQ})] = 0\,,
\\
\\
\displaystyle N: \, \, \, \,  \frac{6 a \dot{a}^2 f_Q}{N} - a^3 N (f - Q f_Q) = 0\,,
\\
\\
\displaystyle Q: \, \, \, \,  Q = \frac{6}{N^2}\left(\frac{\dot{a}^2}{a^2}\right).
\end{cases}
\end{equation}
The third equation of the system is the Lagrange multiplier, namely the cosmological expression of the non-metricity scalar. 
Notice that, by combining the above three equations  with the choice $N=1$, we get: 
\begin{eqnarray}
&& 6 f_Q H^2 - \frac{1}{2} f = 0\,,   
\\
&& 12 f_{QQ} H^2 + f_Q = 0\,,
\end{eqnarray}
which are exactly the field equations coming from the variational principle~\cite{Jimenez:2019ovq} in vacuum. 
The above system of equations, with function $f(Q)$ given in Eq.~\eqref{solf(Q)1}, admits an exponential solution for $a(t)$, so that the scale factor and the non-metricity scalar are:
\begin{equation}
a(t) = a_0 e^{\sqrt{- \frac{\rho_c}{3 \kappa}} \, t}\,, \quad\quad Q(t) = -2 \frac{\rho_c}{\kappa}\,.
\label{deSitter}
\end{equation}
Being $\rho_c/(3 \kappa) > 0$, bouncing cosmological solutions occur, as expected. 

In order to find the Hamiltonian related to the Lagrangian \eqref{solf(Q)1}, we notice that the only conjugate momentum that can be defined is
\begin{equation}
\pi_a = \frac{\partial \Lagr}{\partial \dot{a}} = - \frac{\kappa^2 \, a \dot{a}}{\rho_c N} \left(\frac{3 \tilde{c}_1 \rho_c}{\kappa^2 \sqrt{Q}}+6 \frac{\rho_c}{\kappa} +2 Q\right).
\end{equation} 
By means of the above momentum, it is possible to write the Lagrangian as a function of $\pi_a$; then, after a straightforward Legendre transformation, the Hamiltonian can be written as
\begin{equation}
\mathcal{H} = -N\left[\frac{a^3}{12} \left(\kappa^2 \frac{Q^2}{\rho_c}- 3\sqrt{Q} \,  \tilde{c}_1\right)+\frac{1}{2a} \left(\frac{\rho_c \,  \sqrt{Q}\,  \pi_a^2}{3  \tilde{c}_1 \rho_c +6  \kappa \rho_c \sqrt{Q}+ 2 \kappa^2 Q^{3/2}}\right)\right].
\end{equation}
Considering the quantization rule in Eq.~\eqref{constraints}, the Wheeler--DeWitt equation reads
\begin{equation}
\frac{a^4}{6} \left(\kappa^2 \frac{Q^2}{\rho_c} -3 \sqrt{Q} \,  \tilde{c}_1\right) \psi(a,Q) - \left(\frac{\rho_c \,  \sqrt{Q}\, }{3  \tilde{c}_1 \rho_c +6  \kappa \rho_c \sqrt{Q}+ 2 \kappa^2 Q^{3/2}}\right) \frac{\partial^2 \psi(a,Q)}{\partial a^2} = 0\,,
\label{WDW}
\end{equation}
where $\psi(a,Q)$ is the Wave Function of the Universe. With the definition
\begin{equation}
\alpha(Q,N) \equiv \frac{6 \rho_c \,  \sqrt{Q}}{\left(\kappa^2 \frac{Q^2}{\rho_c} -3 \sqrt{Q} \,  \tilde{c}_1\right)\left(3  \tilde{c}_1 \rho_c +6  \kappa \rho_c \sqrt{Q}+ 2 \kappa^2 Q^{3/2}\right)}\,, 
\end{equation}
 Eq.~\eqref{WDW} takes the form
\begin{equation}
a^4 \psi(a,Q) - \alpha(Q,N) \frac{\partial^2 \psi(a,Q)}{\partial a^2} = 0\,,
\end{equation}
whose solution is a linear combination of first-kind Bessel functions of the form
\begin{equation}
\psi(\alpha, a) = \frac{1}{\alpha^{1/12}} \left[C_2 \sqrt{a} \, \text{J}_{-\frac{1}{6}}\left(\frac{a^3}{3\sqrt{\alpha}} \right) \Gamma \left( \frac{5}{6}\right) + C_3 \sqrt{a} \, \text{J}_{\frac{1}{6}}\left(\frac{a^3}{3\sqrt{\alpha}} \right) \Gamma \left( \frac{7}{6}\right)\right].
\end{equation}
Here $C_2$ and $C_3$ are complex coefficients, while J and $\Gamma$ stand for the Bessel function of the first kind and the Euler gamma function respectively. This latter function can be included in the complex coefficients, so that the Wave Function becomes
\begin{equation}
\psi(\alpha, a) =  \frac{\sqrt{a}}{\alpha^{1/12}} \left[ C_2 \, \text{J}_{-\frac{1}{6}}\left(\frac{a^3}{3\sqrt{\alpha}} \right) + C_3  \, \text{J}_{\frac{1}{6}}\left(\frac{a^3}{3\sqrt{\alpha}} \right)\right].
\end{equation}
Given the fractional index of the Bessel function, $\psi(\alpha,a)$ is a multivalued function whose integral representation can be obtained by means of the Schl\"afli integrals, according to which it can be written as
\begin{eqnarray}
\psi(\alpha, a) = \frac{\sqrt{a}}{\pi \alpha^{1/12}} &\times & \left\{ C_2 \int_{0}^{\pi} \cos \left[- \frac{x}{6} - \left(\frac{a^3}{3\sqrt{\alpha}} \right) \sin x\right] d x+ \frac{C_2}{2} \int_{0}^{\infty} e^{-\left(\frac{a^3}{3\sqrt{\alpha}} \right) \sinh x+ \frac{1}{6} x} d x  +\right. \nonumber 
\\
&+& \left. C_3 \int_{0}^{\pi} \cos \left[\frac{x}{6} - \left(\frac{a^3}{3\sqrt{\alpha}} \right) \sin x\right] d x- \frac{C_3}{2} \int_{0}^{\infty} e^{-\left(\frac{a^3}{3\sqrt{\alpha}} \right) \sinh x- \frac{1}{6} x} d x\right\}.
\label{WFC2C3}
\end{eqnarray}
In order to plot the qualitative trend of the Wave Function, we set $C_2 = C_3 \equiv c$, so that the above expression can be simplified to
\begin{equation}
\psi(\alpha, a) = \frac{c \,\sqrt{a}}{\pi \alpha^{1/12}} \left\{\int_{0}^{\pi} 2 \cos \left[\left(\frac{a^3}{3\sqrt{\alpha}} \right) \sin x\right] \cos\left(\frac{x}{6} \right) d x+ \int_{0}^{\infty} e^{-\left(\frac{a^3}{3\sqrt{\alpha}} \right) \sinh x} \sinh \left(\frac{1}{6} x\right) d x \right\}.
\label{WF}
\end{equation}
Notice that the imposition on the constants $C_2$ and $C_3$ has no implications on the qualitative trend of the Wave Function, as well as the value of $c$. The arbitrary choice $C_2 = C_3 = c$ permits to write Eq.~\eqref{WFC2C3} in a more compact form, but the qualitative behavior of the Wave Function turns out to be the same regardless of the value of $C_2$ and $C_3$.

In Fig.~1, we report two qualitative graphs of the Wave Function in different ranges.
\\
\begin{figure}
\label{Fig}
\includegraphics[width=.90\textwidth]{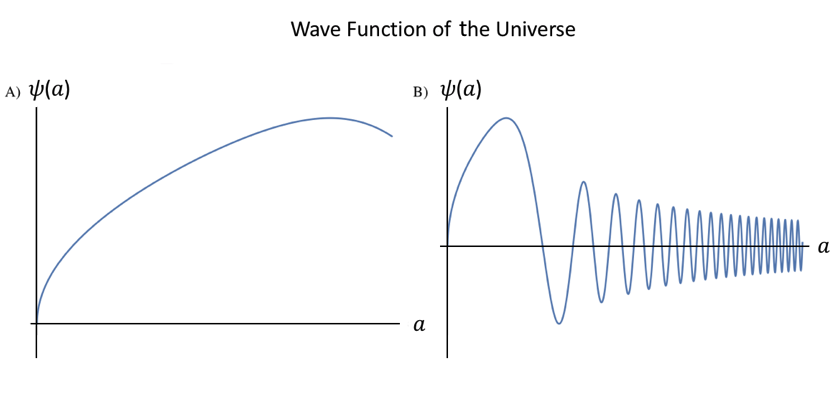}
\caption{Behavior of the Wave Function in two different ranges: in Plot A) small values of the scale factor are considered while Plot B) shows larger values of $a(t)$.}
\end{figure}

The above Wave Function comes from a linear combination of Bessel functions of the first kind and admits as analytic expression only the integral representation. The trend of $\psi$ as a function of the scale factor, shows that the Wave Function increases the oscillation frequency as $a$ increases its value. In the late-time, therefore, the Hartle Criterion holds and  classical universes can be recovered, getting an asymptotically perfect oscillation. 

\section{Conclusions} \label{concl}

Nowadays GR still remains the best candidate to describe the gravitational interaction in the classical regime, despite some shortcomings related to IR and UV behaviors. Nevertheless, during the last decades, several modified theories of gravity have been proposed in order to solve the above issues, mainly related to very small and very large scales. Here we focused on the extension of STEGR, in view of searching for bouncing cosmologies. 

Specifically, we studied the modified STEGR action depending on the function $f(Q)$. With the aim to obtain the FLRW dynamics  provided by LQC, we reduced the field equations applying the well established technique of order reduction. As a side result, we selected \emph{a-posteriori} the form of $f(Q)$ function. 

By means of the order reduction, we found a solution which can be perturbatively related to GR, and the $f(Q)$ function, leading to bouncing cosmology, as prescribed by LQC, is given by Eq.~\eqref{solf(Q)1}.
Among all the possible shapes of the function $f(Q)$, we found the only one presenting cosmological bouncing solutions avoiding the Big Bang initial singularity. 

After that, adopting  the Lagrange Multipliers method, we found the point-like cosmological Lagrangian for $f(Q)$ and  solved the corresponding equations of motion. We found exponential solutions for the scale factor with complex exponent, in accordance to the procedure of order reduction. The approach has been formulated in the minisuperspace containing the variables $a$ and $Q$, considered as independent fields. 

We finally developed the Hamiltonian formalism, with the aim to find the Wheeler--DeWitt equation and the Wave Function of the Universe. The latter turns out to be peaked in the late-time, in agreement with the prescriptions of the Hartle Criterion. This means that for large values of the scale factor, we recover observable universes with the corresponding classical trajectories. The Wave Function can be written as a linear combination of first-kind Bessel functions which are asymptotically periodic, as showed in the plots of Fig.~1. Even though the interpretation of the Wave Function is not entirely clear, from its behavior it is possible to infer important features regarding the early Universe, as well as the transition from a quantum to a classical behavior. 

In future works, we aim to investigate more general functions containing invariants of the non-metricity tensor $Q_{\alpha \beta \gamma}$, selecting the functional form of the gravitational action by order reduction of the field equations and finding the corresponding Wave Function of the Universe.  

\section*{Acknowledgments}

The Authors acknowledge the support of {\it Istituto Nazionale di Fisica Nucleare} (INFN) ({\it iniziative specifiche} GINGER, MOONLIGHT2, QGSKY, and TEONGRAV). This paper is based upon work from COST action CA15117 (CANTATA), COST Action CA16104 (GWverse), and COST action CA18108 (QG-MM), supported by COST (European Cooperation in Science and Technology).

\appendix

\begin{appendix}

\section{Hamiltonian and ADM formalism of General Relativity} \label{HAM}

A quantum description of gravity needs to be developed under the Hamiltonian point of view, as a first step. In this way, in analogy to Quantum Mechanics, one can impose the commutation relations to the quantized Hamiltonian and find the Wave Function of the Universe. It cannot be intended as a standard Wave Function with the same meaning as any other Quantum Field Theory. The main reason is due to the standard probabilistic interpretation of the Wave Function as a probability amplitude, whose squared modulus integrated over the space provides the probability to get a certain configuration. Such an interpretation requires many copies of the same system to make sense, otherwise the concept of probability itself stops being valid. This cannot be applied to gravity and cosmology, since we do not have a final theory of Quantum Gravity and a self-consistent interpretation of probability for the spacetime.  Nevertheless, although the meaning of the Wave Function is still unclear, many interpretations have been given over the years. For instance, according to the so called \emph{Many World Interpretation}, the Wave Function comes from quantum measurements that are simultaneously realized in different universes without, therefore, showing any collapse of the Wave Function as in standard Quantum Mechanics~\cite{Bousso:2011up}. Another interpretation was provided by Hawking, according to whom the Wave Function is supposed to be related to the probability for the early Universe to develop towards our classical Universe~\cite{Vilenkin:1988yd, Hartle:1983ai, Hawking:1983hj}. 

In this scheme of interpretation, J.~B.~Hartle proposed a criterion to gain information from the Wave Function, based on its trend in the late-time. Specifically, according to the Hartle Criterion, the Wave Function must have an oscillating behavior in the classically permitted area, namely whereas it describes our classical Universe~\cite{Hartle:1983ai}. In principle, thanks to the Hartle Criterion and in the Wentzel--Kramers--Brillouin approximation, it is possible to write the Wave Function in terms of the classical action $S_0$ as $\psi \sim e^{i S_0}$. In this way, thanks to the Hamilton--Jacobi equations, we recover the same equations of motion provided by the cosmological Lagrangian and, therefore, the same trajectories. These results make Quantum Cosmology an important connection point between classical and quantum gravity; while waiting for a complete theory of Quantum Gravity, the particular application to cosmology represents a sort of interpretative model capable of reducing the infinite-dimensional superspace coming from the ADM formalism to minisuperspaces whose the equations of motion can be interpreted and, eventually, integrated. However, Quantum Cosmology does not aim to solve the remaining problems of GR due to the UV and IR quantum corrections, as well as the renormalizability problem or the lack of a description under the Yang--Mills formalism. Therefore, the canonical quantization of gravity is not a complete theory, but only aims to solve parts of the high-energy issues arising in standard GR. Without the claim of completeness, let us discuss the basic foundations of the ADM formalism, whose main features can be found in Refs.~\cite{Hawking:1995fd, Kuchar:1976yw, Ashtekar:1987gu}.

In the general form, the Einstein--Hilbert action must include the extrinsic curvature tensor of the three-dimensional spatial surface $K_{ij}$~\footnote{Latin indices denote the four-dimenional flat spacetime, with exception for middle Latin indices, which denote the three-dimensional space; Greek indices denote the four-dimensional curved spacetime.}, the cosmological constant $\Lambda$ and the scalar curvature. It reads as~\cite{DeWitt:1967yk,Thiemann:2007zz}:
\begin{equation}
S = \frac{\kappa}{2} \int_V \sqrt{-g}[R - 2 \Lambda] d^4x + \int_{\partial V} \sqrt{h} K \; dx^3\,.
\label{ADMAction}
\end{equation}
Here $V$ represents the manifold considered and $\partial V$ the three-dimensional spatial surface, while the scalar $K$ is  defined through the extrinsic curvature tensor $K_{ij}$ as $ K = h^{ij} K_{ij}$, where $h_{ij}$ is the spatial metric. Dealing with the three-dimensional surface is important in the view of the (3+1) decomposition of the metric $g_{\mu \nu}$, so that the spatial coordinates account for the dynamical degrees of freedom evolving in the time-line. Specifically, the metric is decomposed so that dynamics is described by the three-dimensional hypersurfaces evolving through the time component. In this way, if $X^{\alpha}$ is a set of coordinates, the transformation $X^\alpha \to X'^\alpha$ can be seen as a family of hypersurfaces with local coordinates $x^i$. Any point of a given hypersurface can be characterized by a three-dimensional tangent vectors basis $X^\alpha_i$, orthonormal to the unitary vector $n^\nu$, labelling the surface. As a consequence, we have the following relations:
\begin{equation}
g_{\mu \nu} X_i^\mu n^\nu = 0\,, \;\;\;\;\;\; g_{\mu \nu} n^\mu n^\nu = -1\,.
\label{vectors}
\end{equation}
The first equation establishes the orthonormality between the vector $n^\nu$ and the set of coordinates $X^\alpha$, while the second is nothing but the parallelism condition between two unitary vectors $n^\nu$ and $n^\mu$. The \emph{Deformation Tensor} can be defined as the time derivative of the coordinates $X^\alpha$:
\begin{equation}
N^\alpha = \dot{X}^\alpha = \partial_0 X^\alpha(x^0, x^i)\,,
\end{equation}
and can be decomposed in the basis of tangent and orthonormal vectors by means of the \emph{Lapse Function} $N^i$ and the \emph{Shift Function} $N$:
\begin{equation}
N^\alpha = N n^\alpha + N^i X_i^\alpha\,.
\end{equation}
With these definitions in mind, the metric tensor can be written in terms of $N$ and $N^i$ as:
\begin{equation}
g_{\mu \nu} = \left(
\begin{matrix}

- (N^2 - N_i N^i) & N_j 
\\
N_j & h_{ij}

\end{matrix}
\right).
\end{equation}
These definitions both with the imposition of vanishing cosmological constant, allow to write the Lagrangian density  in Eq.~\eqref{ADMAction} as: 
\begin{equation}
\mathscr{L}=\frac{\kappa}{2} \sqrt{h} N\left(K^{i j} K_{i j}-K^{2}+{ }^{(3)} R\right) + t. d. \,,
\label{lagradens}
\end{equation}
where $^{(3)}R$ stands for the intrinsic three-dimensional curvature and $h$ is the determinant of the three-dimensional metric $h_{ij}$. Given the Lagrangian of the theory and considering that the dynamical degrees of freedom are $N$, $N^i$ and $h_{ij}$, the conjugate momenta can be written as:
\begin{eqnarray}
&& \pi \equiv \frac{\delta \mathscr{L}}{\delta \dot{N}}=0\,, \quad \pi^{i} \equiv \frac{\delta \mathscr{L}}{\delta \dot{N}_{i}}=0\,, \nonumber \\ 
&& \pi^{i j} \equiv \frac{\delta \mathscr{L}}{\delta \dot{h}_{i j}}=\frac{\kappa \sqrt{h}}{2}\left(K h^{i j}-K^{i j}\right),
\label{conjmom}
\end{eqnarray}
so that, by Legendre transforming the Lagrangian~\eqref{lagradens}, the Hamiltonian density turns out to be 
\begin{equation}
\mathscr{H}=\pi^{i j} \dot{h}_{i j}-\mathscr{L}\,,
\end{equation}
satisfying the constraints 
\begin{equation}
\begin{cases}
\displaystyle \dot{\pi} = -\{\mathcal{H}, \pi \} = \frac{\delta \mathcal{H}}{\delta N} = 0\,,
\\
\displaystyle \dot{\pi}^i = -\{\mathcal{H}, \pi^i \} =  \frac{\delta \mathcal{H}}{\delta N_i} = 0\,,
\end{cases} 
\label{vectors2}
\end{equation}
where $\mathcal{H} = \int \mathscr{H} \, d^3 x$\,.

As usual method to quantize the theory, the first step consists in transforming the dynamical variables into operators and the Poisson brackets in commutators. Therefore, by rewriting the momenta definitions \eqref{conjmom} as
\begin{equation}
\hat{\pi}=-i \frac{\delta}{\delta N}\,, \quad \hat{\pi}^{i}=-i \frac{\delta}{\delta N_{i}}\,, \quad \hat{\pi}^{i j}=-i \frac{\delta}{\delta h_{i j}}\,,
\end{equation}
the following commutation relations hold:
\begin{equation}
\begin{cases}
[\hat{h}_{ij}(x), \hat{\pi}^{kl} (x')] = i \; \delta^{kl}_{ij} \; \delta^3 (x-x')\,,
\\
\delta^{kl}_{ij} = \frac{1}{2} (\delta_i^k \delta_j^l + \delta_i^l \delta_j^k)\,,
\\
[\hat{h}_{ij}, \hat{h}_{kl}] = 0\,,
\\
[\hat{\pi}^{ij}, \hat{\pi}^{kl} ] = 0\,.
\end{cases}
\end{equation}
Finally, the first relation of Eq.~\eqref{vectors2}, in the canonical quantization scheme, becomes
\begin{equation}
\hat{{\cal{H}}}| \psi> = 0\,.
\end{equation}
The above equations, after considering the form of the Hamiltonian as a function of dynamical variables and momenta, leads to a Schroedinger-like equation of the form
\begin{equation}
\left(\nabla^2 - \frac{\kappa^2}{4} \sqrt{h} \; ^{(3)} R \right)|\psi> = 0\,,
\label{constraints}
\end{equation}
called the Wheeler--DeWitt equation. In Eq.~\eqref{constraints},  $\psi$ is the Wave Function of the Universe, depending on the spatial metric $h_{ij}$ and describing the evolution of the gravitational field. 
The operator $\nabla^2$, is defined as
\begin{equation}
\nabla^2 = \frac{1}{\sqrt{h}}\left(h_{i k} h_{j l}+h_{i l} h_{j k}-h_{i j} h_{k l}\right) \frac{\delta}{\delta h_{ij}}\frac{\delta}{\delta h_{kl}}\,.
\end{equation}

In non-relativistic Quantum Mechanics, the scalar product $ \int \psi^* \psi \; dx^3$ is everywhere positive, allowing   the definition of an infinite-dimensional Hilbert space. The main problem related to the Wave Function of the Universe is that it is no possible to define an everywhere positive scalar product due to the hyperbolic nature of Eq.\eqref{constraints} and, therefore, to assign a probabilistic meaning to the Wave Function. 

Nevertheless, the Wave Function may represent an important quantity capable of giving information about the early stages of the Universe on the one hand, and of explaining the nowadays evolution on the other hand. Regarding the latter point, the oscillating Wave Function in the minisuperspaces allows to recover the Hartle Criterion. 
\end{appendix}

\end{document}